\documentclass[runningheads]{llncs}
\usepackage{graphicx}
\usepackage{makeidx}  
\usepackage{cite} 
\usepackage{hyperref}
\usepackage[misc]{ifsym}
\usepackage[utf8]{inputenc} 
\usepackage{graphicx}
\usepackage{array}
\usepackage{multirow}
\usepackage{multicol}
\begin{document}
\title{Transferring Knowledge from High-Quality to Low-Quality MRI for Adult Glioma Diagnosis}
%
\author{Yanguang Zhao\inst{1,2} 
\thanks{This work is done when Y. Zhao and Z. Zhang are research interns at CUHK.}
\and Long Bai\inst{1,2}
\thanks{Project Lead.}
\and Zhaoxi Zhang\inst{1,2~\star}
\and Yanan Wu\inst{3}
\and Mobarakol Islam\inst{4} 
\and Hongliang Ren\inst{1,2}
\thanks{Corresponding Author.}}

\authorrunning{Y. Zhao et al.}
%
\institute{Department of Electronic Engineering, The Chinese University of Hong Kong (CUHK), Hong Kong, China \and CUHK Shenzhen Research Institute, Shenzhen, China \and School of Health Management, China Medical University, Shenyang, China \and UCL Hawkes Institute, University College London, London, UK \\
\email{yanguangzhaozyg@gmail.com, b.long@link.cuhk.edu.hk, jessezhaoxizhang@gmail.com, wuyanan.cmu@vip.163.com, mobarakol.islam@ucl.ac.uk, hlren@cuhk.edu.hk}
}
\maketitle

\begin{abstract}
Glioma, a common and deadly brain tumor, requires early diagnosis for improved prognosis. However, low-quality Magnetic Resonance Imaging (MRI) technology in Sub-Saharan Africa (SSA) hinders accurate diagnosis. This paper presents our work in the BraTS Challenge on SSA Adult Glioma. We adopt the model from the BraTS-GLI 2021 winning solution and utilize it with three training strategies: (1) initially training on the BraTS-GLI 2021 dataset with fine-tuning on the BraTS-Africa dataset, (2) training solely on the BraTS-Africa dataset, and (3) training solely on the BraTS-Africa dataset with 2x super-resolution enhancement. Results show that initial training on the BraTS-GLI 2021 dataset followed by fine-tuning on the BraTS-Africa dataset has yielded the best results. This suggests the importance of high-quality datasets in providing prior knowledge during training. Our top-performing model achieves Dice scores of 0.882, 0.840, and 0.926, and Hausdorff Distance (95\%) scores of 15.324, 37.518, and 13.971 for enhancing tumor, tumor core, and whole tumor, respectively, in the validation phase. In the final phase of the competition, our approach successfully secured second place overall, reflecting the strength and effectiveness of our model and training strategies. Our approach provides insights into improving glioma diagnosis in SSA, showing the potential of deep learning in resource-limited settings and the importance of transfer learning from high-quality datasets.
\end{abstract}


\section{Introduction}

Glioma, a type of tumor originating from the glial cells within the brain, is one of the most common primary brain tumors. It is particularly notable as one of the most prevalent malignant tumors in the brain and central nervous system in adults. Glioma can cause a wide range of symptoms including headaches, motor dysfunction, visual field deficits, seizures, and so on. They are also associated with a high mortality rate.
Approximately 80\% of individuals with glioma die within two years of diagnosis~\cite{poon2020longer}.

Glioma is classified into four grades (I-IV) by the World Health Organization (WHO), with higher grades indicating more aggressive tumors.
In the clinical diagnosis of glioma, imaging diagnostic techniques such as Computed Tomography (CT) and Magnetic Resonance Imaging (MRI) play a crucial role in assisting doctors with localization diagnosis. These methods effectively help doctors ascertain tumor size and extent, and identify important surrounding brain structures, such as major arteries and nerves. 
Currently, the pathogenesis of glioma remains unclear. Therefore, early diagnosis is crucial in glioma treatment, as it allows for timely intervention, potentially at lower grades, which can significantly improve treatment outcomes and patients' prognosis. Moreover, early detection may increase the likelihood of more complete surgical resection and provide a wider range of treatment options.  
This underscores the potential of applying advanced machine learning algorithms to neuroimaging data for more accurate and efficient glioma detection, which could significantly improve early diagnosis and treatment outcomes.

Over the past decades, with the advancement of deep learning methods~\cite{wu2023transformer, wu2023two, bai2023surgical, islam2018class}, numerous initiatives and challenges have been dedicated to harnessing the potential of machine learning for medical image analysis~\cite{wang2023domain, wang2023rethinking, yu2024sam}, yielding increasingly promising outcomes~\cite{wu2021research}. These endeavors have resulted in significant improvements in the accuracy of glioma detection and classification.
Among these challenges, The Brain Tumor Segmentation (BraTS) challenge~\cite{menze2014multimodal,bakas2018identifying,bakas2017segmentation,bakas10segmentation,baid2021rsna} is a well-known annual competition in the medical imaging community. Initiated in 2012, BraTS aims to find state-of-the-art methods for the segmentation of brain tumors in multimodal MRI scans. The challenge provides a large dataset of brain MRI scans from glioma patients, along with expert-annotated ground truth segmentations. Participants are tasked with developing algorithms to automatically segment different tumor sub-regions.
Numerous outstanding algorithms have emerged in the challenge. 
In 2021, the KAIST MRI lab team from South Korea improved upon the nnUNet framework~\cite{isensee2021nnu} by implementing a larger network, replacing batch normalization with group normalization, and incorporating Axial Attention in the decoder~\cite{luu2021extending}. 
The following year, Zimmerer et al. enhanced model prediction performance by integrating three distinct frameworks: DeepSeg~\cite{zeineldin2020deepseg}, nnUNet~\cite{isensee2021nnu}, and DeepSCAN~\cite{mckinley2019ensembles}, and utilizing an ensemble of these models to improve segmentation results. In 2023, Kori et al. focused on data augmentation, introducing GANs and registration techniques to substantially increase the available training samples. They validated the effectiveness of this approach across three different models~\cite{ferreira2024we}.

\begin{figure}
    \centering
    \includegraphics[width=0.75\linewidth]{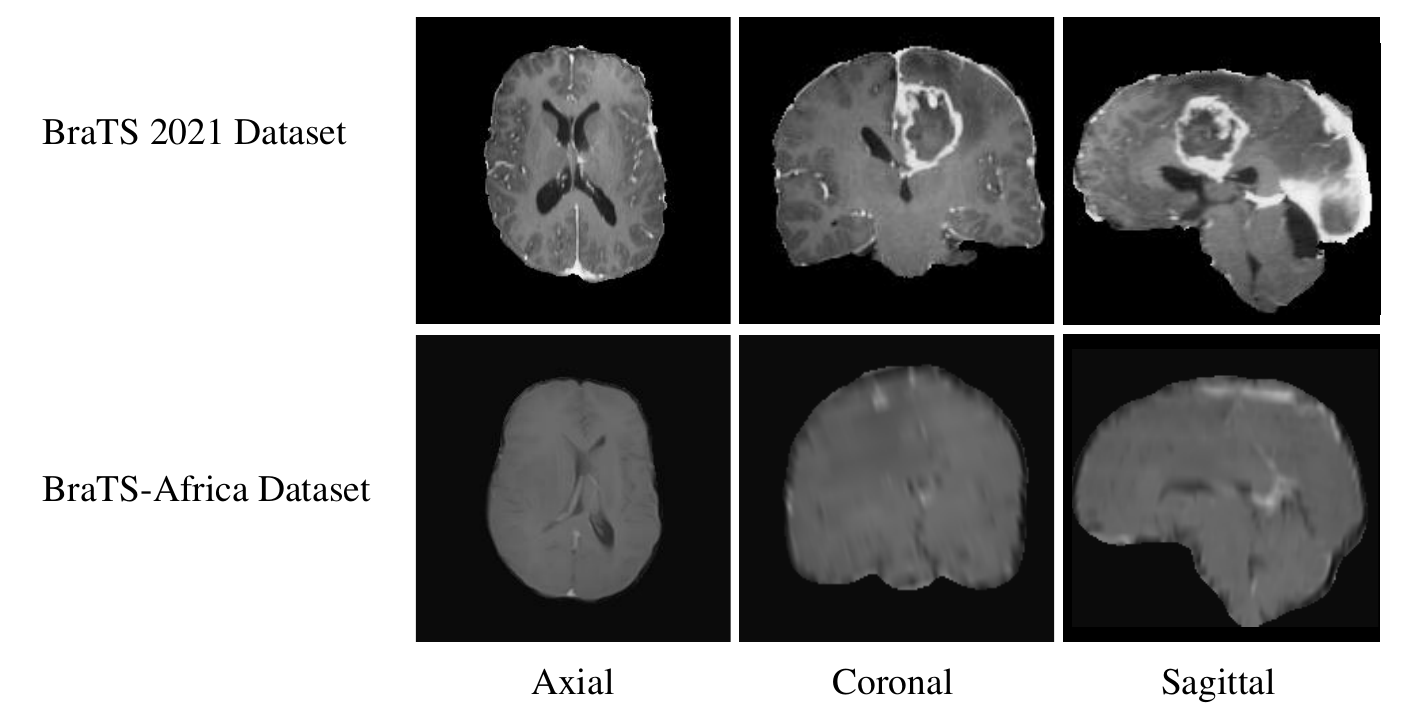}
    \caption{Representative slices respectively in axial, coronal, and sagittal views from two datasets. The upper slices are from the high-quality RSNA-ASNR-MICCAI BraTS-GLI 2021 dataset, and the bottom slices are from the low-quality BraTS-Africa dataset.}
    \label{fig:1}
\end{figure}

Despite advancements in diagnostic and treatment technologies, brain glioma mortality rates show a stark contrast between developed nations and low- and middle-income countries (LMICs). While developed countries have witnessed a decline in glioma-related deaths, LMICs continue to struggle with high mortality rates. This disparity is particularly evident in Sub-Saharan Africa (SSA), where glioma mortality rates have increased by approximately 25\%, contrasting sharply with the 30\% decrease observed in the Global North~\cite{patel2019global}. 
Accurate diagnosis and management of brain gliomas heavily rely on high-quality MRI for precise localization and treatment follow-up. However, as depicted in Fig.~\ref{fig:1}, Sub-Saharan Africa's outdated and low-quality MRI technology often produces images with poor contrast and resolution, significantly hindering accurate diagnosis and effective treatment planning. Furthermore, the performance of existing glioma segmentation algorithms, primarily developed and tested on high-quality imaging data, remains uncertain when applied to these suboptimal images.
Aiming to address the unique challenges of brain tumor diagnosis in Sub-Saharan Africa, the annual BraTS-Africa Challenge~\cite{adewole2023braintumorsegmentationbrats} has been inaugurated in 2023. Supported by CAMERA and the Lacuna Fund, this initiative provides labeled brain MRI glioma datasets from African imaging centers for the first time. The challenge seeks to evaluate the effectiveness of state-of-the-art machine learning methods in the context of Sub-Saharan Africa.

For our entry to the challenge, we design different strategies for dataset utilization and attempt to introduce super-resolution techniques into the glioma segmentation task for Sub-Saharan Africa, aiming to address the low-resolution issue of Sub-Saharan Africa data through super-resolution processing. During our research, we adopt the model from the 2021 winning solution and employ three training data strategies for model training: (1) initially training on the BraTS-GLI 2021 dataset followed by fine-tuning on the BraTS-Africa dataset, (2) training only on the BraTS-Africa dataset, and (3) training on the BraTS-Africa dataset with 2x super-resolution enhancement.

\section{Methods}

\subsection{Data}
In our research, we utilize two datasets: the RSNA-ASNR-MICCAI BraTS-GLI 2021 dataset~\cite{baid2021rsna,menze2014multimodal,bakas2017advancing,karargyris2023federated} and the BraTS-Africa Challenge dataset~\cite{adewole2023braintumorsegmentationbrats,karargyris2023federated}. The former dataset provides multi-institutional routine clinically-acquired multi-parametric MRI (mpMRI) scans of glioma, including 1,251 cases with segmentation labels as the training data. The latter dataset provides mpMRI scans of pre-operative glioma in adult African patients, encompassing both low-grade glioma (LGG) and glioblastoma/high-grade glioma (GBM/HGG). This dataset comprises 60 cases with segmentation labels for training, as well as 35 cases without providing ground truth annotations, which are designated for the validation phase of the challenge. The mpMRI scans are image volumes of T1-weighted (T1), post-contrast T1-weighted (T1Gd), T2-weighted (T2), and T2 Fluid Attenuated Inversion Recovery (T2-FLAIR) MRI. Board-certified radiologists with experience in neuro-oncology annotated these images, delineating the enhancing tumor (ET), non-enhancing tumor core (NETC), and surrounding non-enhancing FLAIR hyperintensity (SNFH).

\subsection{Brain Tumor Segmentation Model}
This study adopts the winning solution from the 2021 Brain Tumor Segmentation Challenge as the brain tumor segmentation model, applying it to the BraTS-Africa Challenge while exploring the use of super-resolution techniques to enhance the training data. Building upon the nnUNet framework~\cite{isensee2021nnu}, the approach incorporates an expanded network architecture and replaces batch normalization with group normalization. During the inference phase, the outputs of the original network and the larger network are ensembled to form the final segmentation output, which is shown in Fig.~\ref{fig:2}.

\subsubsection{Baseline nnU-Net}
The foundation of this approach is the nnU-Net model, a powerful baseline for brain tumor segmentation. This 3D U-Net variant processes $128\times128\times128$ voxel patches through an encoder-decoder structure with skip connections. The architecture features a five-level encoder using strided convolutions for downsampling, paired with a symmetric decoder employing transpose convolutions for upsampling. Key elements include Leaky ReLU activation and batch normalization following convolutions. The model adopts a region-based training strategy~\cite{myronenko20193d,isensee2021nnu,zhao2020bag}, predicting overlapping tumor regions, and incorporates deep supervision through additional sigmoid outputs at multiple resolutions. This baseline design has proven effective in capturing complex spatial relationships in medical imaging data.

\subsubsection{Expanded Network and Group Normalization}
Two significant modifications are 1implemented to capitalize on increased training data availability and enhance model capacity. First, the network is expanded asymmetrically, doubling the filter count in the encoder while maintaining the decoder's original configuration. This expansion increases the maximum filter count to 512, enabling the model to capture more intricate features. Second, batch normalization is replaced with group normalization~\cite{wu2018group}, setting the group count to 32 for optimal performance. This adaptation has been proved particularly beneficial for processing larger 3D volumes with limited batch sizes, significantly improving training stability and overall model performance~\cite{myronenko20193d,jiang2020two}. These enhancements allow the model to better handle data variability and extract more nuanced features from complex 3D medical images.
\begin{figure}[t]
    \centering
    \includegraphics[width=1\linewidth]{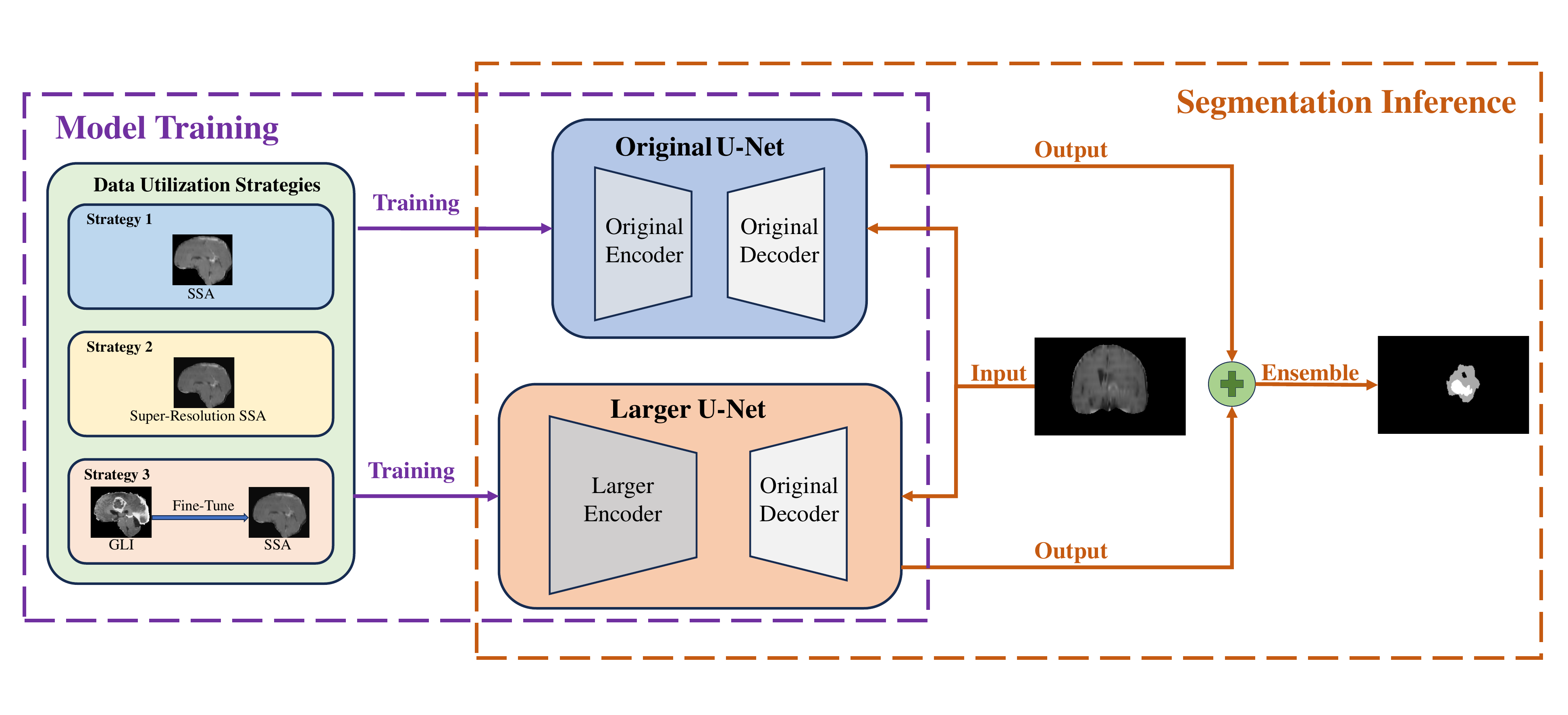}
    \caption{The process of model training and segmentation inference. During the model training process, three data usage strategies are employed to separately train the original network and the larger network. In the inference phase, the outputs of the two networks are ensembled to obtain the final segmentation.}
    \label{fig:2}
\end{figure}
\subsection{Super-Resolution}
\begin{figure}[t]
    \centering
    \includegraphics[width=0.8\linewidth]{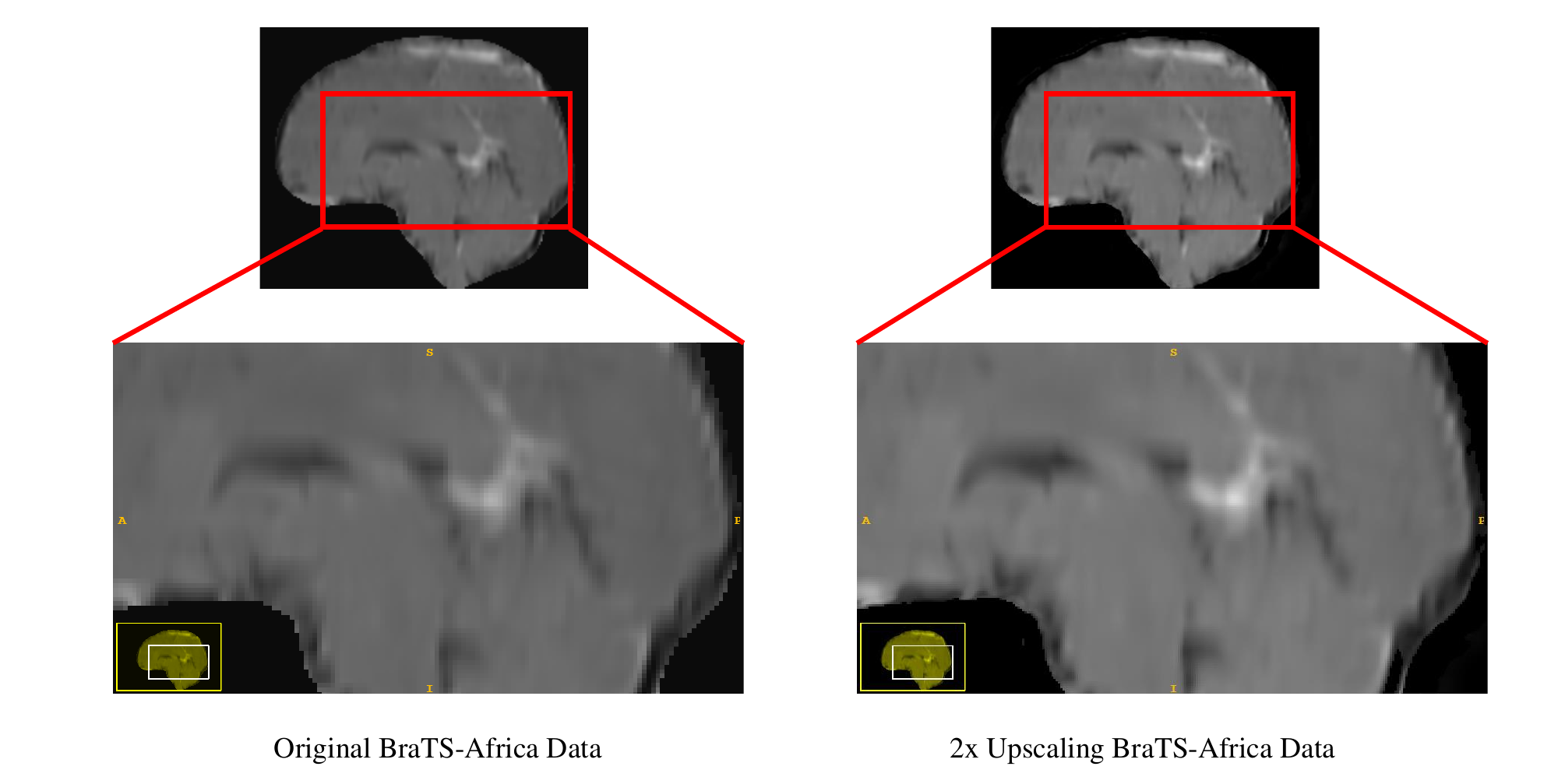}
    \caption{Sagittal slices of the same subject before and after applying super-resolution techniques. The data after enhancement with super-resolution techniques appears smoother, with a more pronounced contrast of internal brain structures.}
    \label{fig:3}
\end{figure}

In our research, we attempt to address the low-resolution issue of the BraTS-Africa Challenge dataset by introducing super-resolution techniques. We employ a 20-layer residual convolutional neural network, which is proposed by Du et al.~\cite{du2020super}, to perform a 2x super-resolution upscaling operation on the data as depicted in Fig.~\ref{fig:3}. The residual convolutional neural network employs an innovative network architecture comprising two key components: the multi-layer convolutional network and the residual learning mechanism. 

\subsubsection{Multi-layer Convolutional Network}
The core of the super-resolution network consists of 19 convolutional layers, each followed by a ReLU activation function~\cite{nair2010rectified} except for the final layer. These convolutional layers are divided into three consecutive blocks, with each block containing 6 convolutional layers, designed to progressively extract and refine image features. To effectively mitigate the vanishing gradient problem in deep networks, the model incorporates short-skip connections, linking blocks formed by every six layers. This design not only facilitates efficient gradient propagation but also preserves crucial image details in deep networks. The final layer employs a single convolutional filter without an activation function to generate the ultimate high-resolution output. 

\subsubsection{Residual Learning Mechanism}
The residual learning mechanism is implemented through a long-skip connection, directly linking the input to the output. This design enables the network to focus on learning the residual mapping, specifically the difference between high-resolution images and interpolated low-resolution images, thereby more effectively recovering high-frequency details. 

\subsection{Training}

\subsubsection{Data Utilization}
For the data utilization strategy, we develop three different approaches as shown in Fig.~\ref{fig:2}: (1) The first approach comprises an initial training of the model utilizing the RSNA-ASNR-MICCAI BraTS-GLI 2021 dataset, subsequently followed by fine-tuning with the BraTS-Africa Challenge dataset ($S_{GLI \rightarrow SSA}$). This approach aims to enable the model to acquire knowledge of brain glioma segmentation from high-quality, extensive data initially, and then adapt to low-quality data tasks through fine-tuning. (2) The second approach consists of complete model training using exclusively the BraTS-Africa Challenge dataset ($S_{SSA}$). This strategy is designed to assess whether the model can effectively learn from and perform optimally on low-quality data alone, thereby potentially excelling in this specific challenge. (3) The third approach implements super-resolution techniques to apply a 2x resolution enhancement to the BraTS-Africa Challenge dataset. Concurrently, resampling is applied to proportionally enlarge the diagnostic files. Subsequently, this augmented dataset is employed for model training ($S_{srSSA}$). The objective of this approach is to evaluate the potential of super-resolution techniques in mitigating the adverse effects of the low resolution of the BraTS-Africa Challenge data.

\subsubsection{Model Training}
For the model training process, we adhere to the training methodology utilized by the winning solution of the 2021 Brain Tumor Segmentation Adult Glioma Challenge. Real-time data augmentation techniques are implemented during the training phase to enhance the model's ability to generalize. These augmentation methods include random rotations and scaling adjustments, elastic deformations, brightness modifications, and gamma value alterations. The optimization process focuses on a combined loss function that incorporates the binary cross-entropy loss and the dice loss. This composite loss was assessed not only at the final full-resolution output but also at intermediate lower-resolution outputs. Importantly, the batch dice loss is implemented in place of sample-wise calculation. This method considers the entire batch as a single unit for loss computation, rather than averaging individual dice scores within the minibatch. The use of batch dice has been demonstrated to stabilize training, particularly by reducing errors in samples with sparse annotations~\cite{isensee2021nnu}. For optimization, the networks utilize stochastic gradient descent, incorporating the Nesterov momentum set at 0.99. The learning rate is initially set to 0.01 and a polynomial decay schedule is followed throughout the training process.

\section{Results}
\begin{figure}[t]
    \centering
    \includegraphics[width=\linewidth]{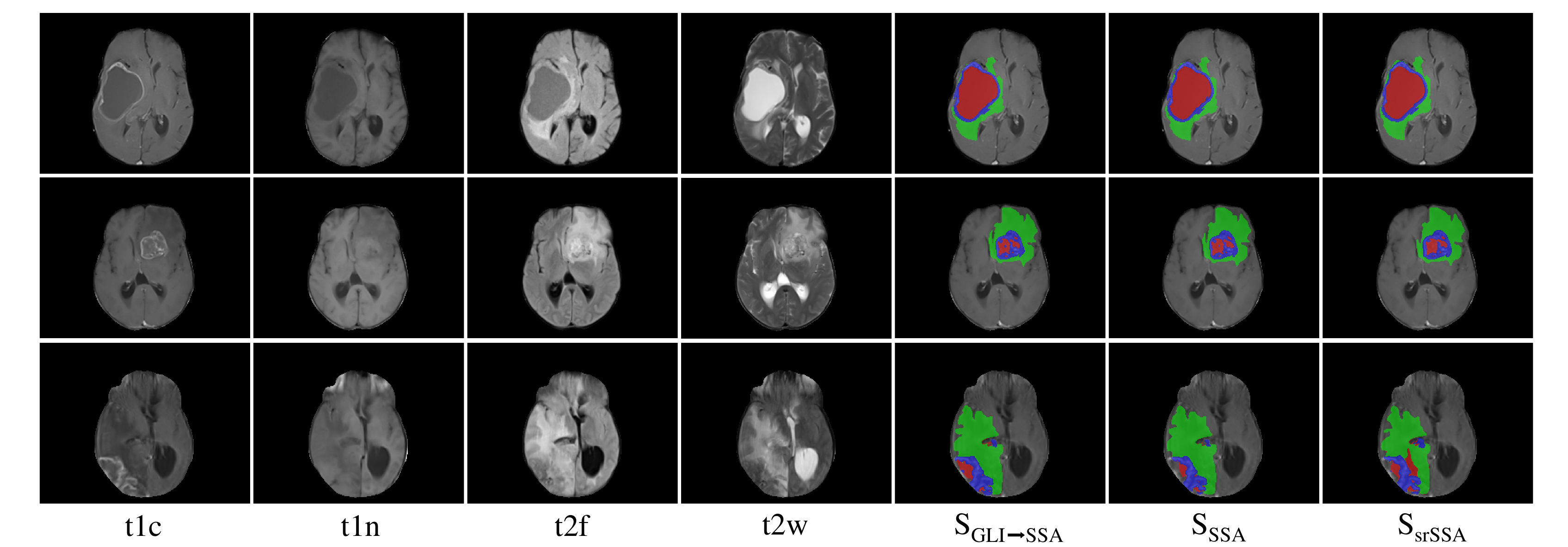}
    \caption{Comparison of segmentation results using the three strategies—$S_{GLI\rightarrow SSA}$, $S_{SSA}$, and $S_{srSSA}$ on selected validation samples. The color coding indicates the non-enhancing tumor core (red), surrounding non-enhancing FLAIR hyperintensity (green), and enhancing tumor (blue).}
    \label{fig:4}
\end{figure}

\begin{figure}
    \centering
    \includegraphics[width=\linewidth]{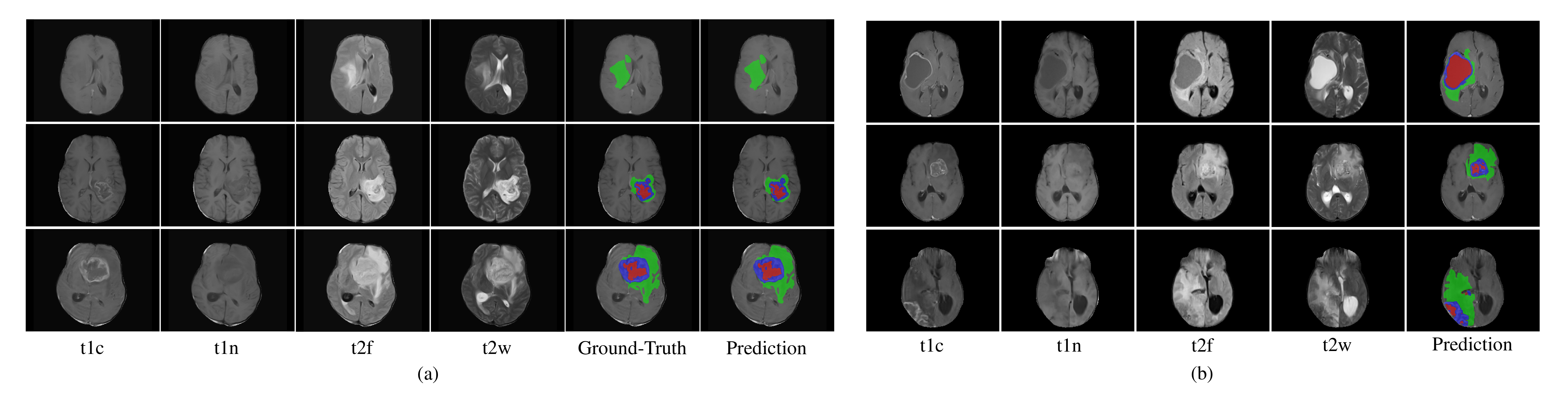}
    \caption{Segmentation results from final submission on selected examples. (a) shows the results on the training data, and (b) on the validation data. The color coding indicates the non-enhancing tumor core (red), surrounding non-enhancing FLAIR hyperintensity (green), and enhancing tumor (blue).}
    \label{fig:5}
\end{figure}

\begin{table}[b]
  \centering
  \caption{The Dice Score and Hausdorff Distance (95\%) results from the validation phase of the BraTS-Africa Challenge using three different approaches: (1) $S_{GLI \rightarrow SSA}$ initially trains on the BraTS-GLI 2021 dataset, subsequently followed by fine-tuning with the BraTS-Africa Challenge dataset. (2) $S_{SSA}$ uses exclusively the BraTS-Africa Challenge dataset. (3) $S_{srSSA}$ implements super-resolution techniques to apply a 2x resolution enhancement to the BraTS-Africa Challenge dataset.}
  \label{table:1}
  \begin{tabular}{c|cccc|cccc}
    \hline
    \multirow{2}{*}{Data Utilization} & \multicolumn{4}{c|}{Dice Score} & \multicolumn{4}{c}{Hausdorff Distance (95\%)}\\
    \cline{2-9}
    & ET & TC & WT & Mean & ET & TC & WT & Mean \\
    \hline
    $S_{GLI\rightarrow SSA}$ & 0.882 & 0.840 & 0.926 & 0.882 & 15.324 & 37.518 & 13.971 & 22.271 \\
    $S_{SSA}$ & 0.844 & 0.838 & 0.898 &  0.860 & 31.484 & 34.849 & 24.742 & 30.358 \\
    $S_{srSSA}$ & 0.822 & 0.831 & 0.886 &  0.846 & 30.920 & 34.464 & 25.275 & 30.219 \\
    \hline
  \end{tabular}
\end{table}
In the validation phase of the BraTS-Africa Challenge, participants are required to evaluate 35 validation samples, using the Dice Similarity Coefficient and Hausdorff Distance (95\%) as assessment metrics~\cite{amod2023bridging}. Table~\ref{table:1} presents the evaluation results for the three training data strategies during the validation phase and Fig.~\ref{fig:4} showcases the comparison of these strategies. The strategy of initially training on the BraTS-GLI 2021 dataset and then fine-tuning on the BraTS-Africa dataset ($S_{GLI\rightarrow SSA}$) yields the best results. It indicates that the rich, high-quality data from the BraTS-GLI 2021 dataset provides the model with excellent prior knowledge for brain glioma segmentation while fine-tuning with the BraTS-Africa dataset enables the model to better handle the low-quality data from Sub-Saharan Africa. The strategy of using the BraTS-Africa dataset ($S_{SSA}$) also achieves good results, suggesting the model from the 2021 winning solution maintains a relatively high level of robustness when dealing with low-quality data. However, the strategy of using the BraTS-Africa dataset upscaled by super-resolution techniques ($S_{srSSA}$) does not lead to the expected performance improvement. It may be due to the loss of initial information caused by the inferior MRI technology in Africa, and the use of super-resolution techniques could not compensate for the negative impact of the loss. In the test phase of the BraTS-Africa Challenge, we choose the model trained with  $S_{GLI\rightarrow SSA}$ as our final submission. Fig.~\ref{fig:5} illustrates selected results from our final submission on the training and validation data. Ultimately, we secure second place in the final ranking, demonstrating the effectiveness of our method.

\section{Conclusion and Discussion}
In this paper, we focus on the BraTS-Africa Challenge, applying three different data training strategies to the model of the BraTS-GLI 2021 winning solution. We achieve second place in the final standings with our best method. Results from the validation phase and the test phase indicate that there are two key factors in addressing the low-quality data issue in the BraTS-Africa Challenge. Firstly, it is crucial to select a high-quality brain tumor segmentation model to ensure that the model can fully employ the data during training to achieve better segmentation results. This is the fundamental basis for ensuring the efficacy of subsequent optimization initiatives. Secondly, adopting a training data strategy that initially trained on the BraTS-GLI 2021 Challenge dataset followed by fine-tuning on the BraTS-Africa dataset helps the model to perform more excellently on the BraTS-Africa Challenge.
Our attempt with super-resolution techniques does not yield the expected ideal results. 
However, we cannot definitively conclude that super-resolution techniques are ineffective in this challenge, as time constraints prevent us from testing other models. 
Furthermore, since super-resolution techniques enlarge the data, it remains uncertain if common preprocessing can handle this and if the same training period allows the model to fully capture the information. 

\begin{credits}
\subsubsection{\ackname} This work was supported by Hong Kong RGC General Research Fund (GRF) 14211420, Collaborative Research Fund (CRF) C4063-18G, NSFC/RGC Joint Research Scheme N\_CUHK420/22; Shenzhen-HK-Macau Technology Research Programme (Type C) STIC Grant 202108233000303; Regional Joint Fund Project 2021B1515120035 (B.02.21.00101) of Guangdong Basic and Applied Research Fund. M. Islam was funded by EPSRC grant [EP/W00805X/1].

\subsubsection{\discintname}
The authors have no competing interests to declare that are relevant to the content of this article.
\end{credits}

%
%
%
\bibliographystyle{splncs04}
\bibliography{mybibliography}

\end{document}